\documentclass[
reprint,
superscriptaddress,
 amsmath,amssymb,
 aps,
]{revtex4-1}

\usepackage{graphicx}
\usepackage{dcolumn}
\usepackage{bm}
\usepackage[mathlines]{lineno}
\usepackage[colorlinks=true]{hyperref}
\usepackage{mathrsfs}
\usepackage[usenames, dvipsnames]{xcolor}
\usepackage{float}
\usepackage{ulem}
\usepackage[utf8]{inputenc}
\usepackage{csquotes}
\newcommand{\be}{\begin{equation}}
\newcommand{\eb}{\end{equation}}

\begin{document}

\title{Charged accelerating black hole in $f(R)$ gravity}

\author{Ming Zhang}
\email{mingzhang@jxnu.edu.cn}
\affiliation{College of Physics and Communication Electronics, Jiangxi Normal University, Nanchang 330022, China}
\affiliation{Department of Physics, Beijing Normal University, Beijing 100875, China}
\author{Robert B. Mann}
\email{rbmann@uwaterloo.ca}
\affiliation{Department of Physics and Astronomy, University of Waterloo,
Waterloo, Ontario N2L 3G1, Canada}
\affiliation{Perimeter Institute for Theoretical Physics, 31 Caroline St. N., Waterloo, Ontario N2L 2Y5, Canada}

\begin{abstract}
We obtain a charged accelerating AdS black hole solution in $f(R)$ gravity  and investigate its thermodynamic behaviour.   We consider low-acceleration black holes that do not have an acceleration horizon and obtain the first law of thermodynamics for them. We further study the parameter space of charged slowly accelerating $f(R)$ AdS black holes before investigating the behaviour of 
the free energy in both the canonical and grand canonical ensembles.  We find a generalization of
the reverse isoperimetric inequality, applicable to black holes in $f(R)$ gravity, that indicates
these black holes can become super-entropic relative to their counterparts in Einstein gravity.
\end{abstract}

\maketitle

\section{Introduction}

The well known vacuum C-metric was introduced in \cite{Weyl:1917gp,levi1918t}, and then investigated in \cite{newman1961new,robinson1962robinson}. Later on, a charged C-metric was discovered \cite{Kinnersley:1970zw}. Mathematically, the C-metric is a solution of the Einstein(-Maxwell) field equations; physically, it can be  interpreted as two uniformly accelerating black holes dragged by forces originating from conical singularities \cite{Griffiths:2006tk}. An accelerating black hole with not only charge but also rotation was raised in \cite{Plebanski:1976gy}.

The C-metric with a negative (positive) cosmological constant has been used as a test background to study gravitationally and electromagnetically radiative fields near timelike (spacelike) infinity \cite{Podolsky:2003gm,Krtous:2003tc}. However, in contrast to the flat or de Sitter (dS) C-metrics, the anti-de Sitter (AdS) C-metric can describe a pair of accelerated black holes only when the acceleration parameter $A$ and the AdS radius $l$ satisfy a certain restriction, originally thought to be  
$A>1/l$ \cite{Dias:2002mi}.  It has recently been shown that this condition must be corrected to be a more stringent one \cite{Anabalon:2018ydc}. If this condition  is not satisfied (which will be the case for sufficiently small acceleration A), then the AdS C-metric  describes an single uniformly accelerating black hole
having an outer (and possibly inner) horizon \cite{Dias:2002mi,Podolsky:2002nk,Anabalon:2018ydc}.

For quite some time the thermodynamics of   accelerating black holes have been poorly understood (and indeed avoided) in the literature, most likely because they generically have two horizons, except for the
slow-acceleration case noted above.  However it has recently been shown that it is possible to make
sense of their thermodynamics in the slow-acceleration case, and a first law for these objects was
derived  \cite{Appels:2016uha}. Furthermore, by allowing the tension related to the conical deficits to vary, thermodynamics of the accelerating black hole was generalized to include a new potential named thermodynamic length \cite{Appels:2017xoe}.  A holographic interpretation of the thermodynamics for a slowly accelerating AdS black holes was subsequently given  \cite{Anabalon:2018ydc},
  and thermodynamic variables were found to be consistently calculated using both conformal and holographic methods. Conserved charges were obtained by choosing a canonical time coordinate and using a covariant phase space method in \cite{Astorino:2016ybm}, as well as checked by the Christodoulou-Ruffini mass formula  \cite{Christodoulou:1972kt}.    An explication of the  thermodynamic parameters of accelerating AdS black holes with charge and rotation was studied in \cite{Anabalon:2018qfv}, and these results were used to obtain  a remarkable snapping-swallowtail phase transition for the charged case at a certain pressure  \cite{Abbasvandi:2018vsh}.  For the rotating case this particular pressure `splits' into three values that have the  effect of admitting reentrant phase transitions obtainable  in two ways, either by varying the temperature at fixed pressure or varying the pressure at fixed temperature \cite{Abbasvandi:2019vfz}.

Until now, all extant C-metrics describing accelerating black holes are solutions of the equations of motion in Einstein's classical gravitational theory. It is of interest to find accelerating black hole solutions in modified gravity theories, such as the renowned $f(R)$ gravity \cite{Sotiriou:2008rp}. A number of  reasons whose origins are in  high-energy physics \cite{Utiyama:1962sn,Stelle:1976gc,Vilkovisky:1992pb} and cosmology \cite{Carroll:2000fy,Weinberg:1988cp} motivate consideration of  higher-order curvature invariants to the gravitational  action. Amongst the various possibilities, $f(R)$ gravity is   an excellent candidate for two reasons. First, it encapsulates the features of the higher-order gravity by, for example, taking a simple form
\begin{equation}
f(R)=\sum_{i}\alpha_i R^i,
\end{equation}
where $R$ is the Ricci scalar and the  $\alpha_{i}$ are coefficients; other higher-order terms such as $R_{\mu\nu}R^{\mu\nu}$  are not considered \cite{Sotiriou:2008rp}. Second, it may be the only
higher-derivative theory that
  avoids the Ostrogradski instability \cite{Woodard:2006nt}.  
  
In this paper, we obtain charged accelerating black hole solutions in $f(R)$ gravity and study their thermodynamics.   The thermodynamics of charged black holes in $f(R)$ gravity have been studied
previously \cite{Mo:2016ndm}, and  phase transitions analogous to the van der Waals liquid-gas system 
 between   small and large black holes were observed to take place.  In our extension to this case we
 find a number of new results.  One is a generalization of the reverse isoperimetric inequality \cite{Cvetic:2010jb} that includes both conical deficits and the effects of $f(R)$ gravity.  We find that $f(R)$ gravity
 can yield both sub-entropic and super-entropic black holes, depending on the magnitude of 
$\eta = 1+ f^\prime(R_0)$, where $R_0$ is value of the Ricci scalar for the accelerating solution. Furthermore, we find that the  parameter space of allowed solutions is either enlarged or diminished, depending on the size of $\eta$. Our investigation of the phase behaviour of these black holes 
indicates that both the expected van der Waals transitions and snapping swallowtail behaviour
observed previously  \cite{Abbasvandi:2018vsh} occur here as well, but we also find new behaviour
in the fixed-potential ensemble, including Hawking-Page-type curves for unstable and stable black holes.
We find that not only the lowest temperature attainable for the black hole but also the highest temperature that the small black hole can reach  increases for increasing $\eta$. Furthermore, both kinds of temperatures have depend on the acceleration of the black hole, as we shall demonstrate.

 We begin by introducing a charged accelerating black hole solution in $f(R)$ gravity and calculating its thermodynamic quantities in Sec. \ref{sectw}. We will analyze  the parameter space and phase transition for the charged slowly accelerating $f(R)$ AdS black hole in Sec. \ref{secth}. The final section is devoted to our summary.

\section{Charged accelerating black hole in $f(R)$ gravity}\label{sectw}
\subsection{Solution}
The action of $f(R)$ gravity containing Maxwell term is
\begin{equation}\label{sou}
\mathcal{I}=\frac{1}{16\pi G}\int d^{4}x\sqrt{-g}\mathcal{L},
\end{equation}
with
\begin{equation}
\mathcal{L}=R+f(R)-F_{ab}F^{ab}.
\end{equation}
Here $R$ is the Ricci scalar, $f(R)$ is an auxiliary function of $R$, $F_{ab}=\nabla_{a}B_{b}-\nabla_{b}B_{a}$, and $B_{a}$ is the gauge potential.
The corresponding equations of motion are
\begin{eqnarray}\label{eos1}
R_{ab}&-&\frac{1}{2}Rg_{ab}-\frac{1}{2}f(R)g_{ab}+R_{ab}f^{\prime}(R)-\nabla_b \nabla_a R f^{\prime\prime}(R)\nonumber\\&+&g_{ab}\nabla_{c}\nabla^{c}Rf^{\prime\prime}(R)\nonumber-\nabla_a R \nabla_b R f^{(3)}(R)\\&+& g_{ab}\nabla_c R\nabla^{c} Rf^{(3)}(R)=2T_{ab}
\end{eqnarray}
and
\be\label{eos2}
\nabla_b \nabla^b B^a-\nabla_b \nabla^a B^b=0,
\eb
with
\be
f^{\prime}(R)=\frac{df(R)}{dR},
\eb
\be
T_{ab}=F_{a}^{~c}F_{bc}-\frac{1}{4}F_{cd}F^{cd}g_{ab}.
\eb
It is easy to check that the electromagnetic field is traceless, as
\begin{equation}
T^{\mu}_{~\mu}=0.
\end{equation}
Supposing that we want to get a maximally symmetry solution, we can know that the Ricci scalar must be constant \cite{Sotiriou:2008rp}. When the curvature scalar is constant, i.e., $R=R_0~(R_{0}\neq 0)$, according to (\ref{eos1}), we can obtain \cite{Moon:2011hq,Sheykhi:2012zz}
\be\label{sef}
R_0 -R_{0} f^{\prime}(R_0) +2 f(R_0)=0.
\eb
Then  (\ref{eos1}) can be re-expressed as
\be\label{eos11}
\eta R_{ab}-\frac{\eta}{4}R_{0}g_{ab}=2T_{ab},
\eb
where we have defined
\begin{equation}\label{fr2}
\eta =1+f^{\prime} (R_0).
\end{equation}

According to the equation of motion (\ref{eos11}), we can obtain a charged accelerating black hole solution
\begin{eqnarray}\label{met}
\begin{aligned}
ds^2=\frac{1}{\Omega^2}&\left[-\frac{N(r)dt^2}{\alpha^2}+\frac{dr^2}{N(r)}\right.\\&\left.+r^2 \left(\frac{d\theta^2}{H(\theta)}+H(\theta)\sin^2\theta \frac{d\phi^2}{K^2}\right)\right],
\end{aligned}
\end{eqnarray}
where
\begin{equation}
\Omega=1+Ar\cos\theta,
\end{equation}
\begin{equation}
N(r)=(1-A^2 r^2)\left(1-\frac{2m}{r}+\frac{q^2}{\eta r^2}\right)-\frac{R_0 r^2}{12},
\end{equation}
\begin{equation}
H(\theta)=1+2mA\cos\theta+\frac{q^2}{\eta}A^2\cos^2 \theta,
\end{equation}
\begin{equation}\label{alpha}
\alpha=\sqrt{\Xi (1+\frac{12 A^2 \Xi}{R_0})},
\end{equation}
\begin{equation}
\Xi=1+\frac{q^2 A^2}{\eta}.
\end{equation}
In the black hole solution, the conformal factor $\Omega$ determines the conformal boundary $r_{b}$ of the black hole and $r_{b}=-1/(A\cos\theta)$. $A,~m,~q$ are individually the acceleration, the mass parameter and the electric parameter of the black hole. $K$ stands for the conical deficits of the black hole on the north and south poles. The parameter $\alpha$ is used to rescale the time coordinate so that we can get a normalized Killing vector at the conformal infinity \cite{Podolsky:2002nk,Anabalon:2018ydc,Anabalon:2018qfv}.  When $A=0$ we recover the charged AdS black hole in $f(R)$ gravity
\cite{Nojiri:2014jqa}.

$R_{0}<0,~R_{0}=0$ and $R_{0}>0$ correspond to asymptotically AdS, flat, and dS accelerating black holes, respectively. In the following, we will only investigate the case $R_{0}<0$. When $A=0,~K=1~R_{0}<0$, the solution  reduces to the $f(R)$ black hole \cite{Moon:2011hq} for which we need $\eta>0$ to assure the existence of inner and outer horizons  \cite{Sheykhi:2012zz}. For consistency, we will only consider the case $\eta>0$ for the accelerating $f(R)$ AdS black hole here. Then we can solve the equation of motion (\ref{eos2}) for the gauge field to obtain the electromagnetic tensor
\begin{equation}
F_{ab}=(dB)_{ab},
\end{equation}
where 
\begin{equation}\label{B-pot}
B_{a}=\frac{1}{\alpha}\left(\frac{q}{r_+}-\frac{q}{r}\right)(dt)_{a},
\end{equation}
and $r_{+}$ is the outer horizon.

When
\begin{equation}
f^{\prime}(R_0)=0,~R_0 =-\frac{12}{l^2},
\end{equation}
the solution is identical to the charged accelerating AdS black hole in Einstein gravity. Hereafter, we will
write   $R_0= -12/l^2$, so that the blackening factor of the black hole can be written as
\begin{equation}\label{mmet}
N(r)=\left(1-\frac{2m}{r}+\frac{q^2}{\eta r^2}\right)(1-A^2 r^2)+\frac{r^2}{l^{2}}
\end{equation}
and we note that $\eta$ and $\l$ are independent parameters.

\subsection{Thermodynamic first law of slowly accelerating $f(R)$ AdS black hole}
We will investigate slowly accelerating $f(R)$ AdS black hole here. The condition of the slow acceleration will be discussed later. The mass of the slowly accelerating AdS black hole can be obtained by conformal prescription \cite{Das:2000cu,Ashtekar:1999jx}. We can choose a conformal factor $\bar{\Omega}=\eta^{-1}l\Omega r^{-1}$ to eliminate the divergence near the conformally infinite boundary.  After a conformal transformation
\begin{equation}
\bar{g}_{ab}=\bar{\Omega}^2 g_{ab}
\end{equation}
on the spacetime metric (\ref{met}), we can obtain the conserved charge related to the Killing vector $\xi^{a}$ as
\begin{equation}
\mathcal{Q}(\xi_{c})=\frac{l}{8\pi}\lim_{\bar{\Omega}\to 0}\frac{l^2}{\bar{\Omega}}{\bar{N}}^{a}{\bar{N}}^{b}\bar{C}^{c}_{~a d b}\xi_{c}d\bar{S}^{d},
\end{equation}
where ${\bar{N}}_{a}$ is a normal covector of the conformal boundary determined by $\bar{\Omega}$, $\bar{C}^{a}_{~b c d}$ is the Weyl tensor of the conformal metric $\bar{g}_{ab}$. After letting $\bar{\Omega}\to 0$ in the conformal metric $\bar{g}_{ab}$, we can obtain the spacelike area element as
\begin{equation}
d\bar{S}_{a}=-\frac{\eta^{2}l^{2}\sin\theta d\theta d\phi}{\alpha K}(dt)_{a}.
\end{equation}
Then the mass $M$ of the slowly accelerating $f(R)$ AdS black hole can be calculated as
\begin{equation}\label{mass}
M=\mathcal{Q}(\partial_t)=\frac{\eta m(1-A^2 l^2 \Xi)}{K\alpha}.
\end{equation}
It should be noted that we have chosen the conformal factor so that the resulting mass of the slowly accelerating $f(R)$ AdS black hole becomes that of the $f(R)$ black hole \cite{Sheykhi:2012zz} calculated by the quasilocal approach \cite{Brown:1992br,Brown:1992bq,Kim:2013zha}.

In contrast to  the slowly accelerating AdS black hole in Einstein gravity, the entropy of the accelerating AdS black hole in $f(R)$ gravity cannot be simply calculated from the area law \cite{Bekenstein:1973ur}; rather the Wald method \cite{Wald:1993nt} must be used. This gives
\begin{equation}\label{accent}
S=-2\pi\oint d^{2}x\sqrt{\hat{h}}\frac{\partial \mathcal{L}}{\partial R_{abcd}}\hat{\epsilon}_{ab}\hat{\epsilon}_{cd}=\frac{\eta\pi r_{+}^{2}}{K(1-A^2 r_+^2 )},
\end{equation}
where $\hat{h}$ is the determinant of the induced metric on the $t=\text{const.}$ and $r=r_{+}$ hypersurface,  and $\epsilon_{ab}$ is a normal bivector which satisfies $\epsilon_{ab}\epsilon^{ab}=-2$.

The temperature of the black hole can be obtained as
\begin{eqnarray}\label{temp}
T&=&\frac{f^{\prime}(r_+)}{4\pi \alpha}\nonumber\\&=&\frac{ l^2 \left(A^2 r_+^2-1\right)^2 \left(q^2-\eta  r_+^2\right)-\eta  r_+^4 \left(3-A^2 r_+^2\right)}{4 l^2 \pi  \alpha  \eta  r_+^3 \left(A^2 r_+^2-1\right)}\nonumber \\&=&\frac{r_+}{2 \pi  \alpha  l^2 \left(1-A^2 r_+^2\right)}+\frac{\left(A^2 r_+^2-1\right) \left(q^2-\eta  m r_+\right)}{2 \pi  \alpha  \eta  r_+^3}.
\end{eqnarray}
Then we can know 
\begin{equation}
TS=\frac{\eta m}{2\alpha K}-\frac{q^2}{2\alpha K r_+}+\frac{\eta r_+^3}{2\alpha l^2 K (1-A^2 r_+^2)^2}.
\end{equation}

The electric charge of the black hole is
\begin{equation}\label{ele}
Q=\frac{1}{4\pi }\lim_{\Omega\to 0}\int\ast F=\frac{1}{4\pi}\int \frac{q}{K}\sin\theta d\theta d\phi =\frac{q}{ K}
\end{equation}
from Gauss' law.  
The conjugate electric potential  is
\begin{align}
\Phi=\frac{1}{4\pi Q \beta}\int_{\partial \mathcal{M}}d^{3}x\sqrt{|h|}n_a F^{ab}B_b=\frac{q}{\alpha r_+}
\label{pot}
\end{align}

where $n_{a}$ is a normal covector on the conformal boundary determined by $\Omega$, $h$ is the determinant of the induced metric on the conformal boundary, $\beta$ is related to the surface gravity $\kappa$ at the event horizon $r_{+}$ by $\beta=2\pi/\kappa$ \cite{Hawking:1995ap}.

Note that although a rescaling of  $q$ 
\begin{equation}
\frac{q^{2}}{\eta}=Q_{f}^{2}
\end{equation}
renders the black  hole solution (\ref{mmet}) the same as that for the accelerated charged black hole in
 Einstein gravity \cite{Anabalon:2018qfv}, this parameter actually does not scale out of the thermodynamics.  The reason is that  that mass \eqref{mass} and entropy \eqref{accent} retain their $\eta$-dependence, and in fact the absence of $\eta$ in \eqref{ele} and \eqref{pot} is essential in satisfying the first law.  We have checked that the gauge potential \eqref{B-pot} can not contain a factor $1/\sqrt{\eta}$, or else \eqref{eos11} can not be satisfied.  Consequently we disagree with the conserved electric charge and conjugate potential previously obtained for  $f(R)$ black holes   \cite{Sheykhi:2012zz,Mo:2016ndm}, where there are respective factors of $1/\sqrt{\eta}$ and $\sqrt{\eta}$.

The pressure of the slowly accelerating $f(R)$ AdS black hole can be defined as \cite{Kastor:2009wy,Dolan:2011xt,Kubiznak:2012wp,Kubiznak:2016qmn,Sheykhi:2012zz,Chen:2013ce,Zhang:2016wek,Mo:2016ndm,Ovgun:2017bgx}
\begin{equation}\label{pressure}
P=\frac{3}{8\pi l^2}.
\end{equation}
Then the volume can be obtained as
\begin{equation}\label{vol}
V=\frac{4\pi \eta}{3K\alpha}\left[\frac{r_+^3}{(1-A^2 r_+^2)^2} +ml^4 A^2\Xi\right]
\end{equation}
by using Smarr relation
\begin{equation}\label{smarr}
M=2(TS-PV)+\Phi Q.
\end{equation}

The conical deficits at the north pole ($\theta_{+}=0$) and south pole ($\theta_{-}=\pi$) are
\begin{equation}
\delta_\pm =2\pi \left(1-\frac{H_{\pm}}{K}\right),
\end{equation}
where
\begin{equation}
H_{\pm}=\Xi\pm 2mA.
\end{equation}
Then the tensions related to the conical deficits on the north and south poles are \cite{Bayona:2010sd,Emparan:1999wa,Dehghani:2001nz}
\begin{eqnarray}
\mu_{\pm}&=&\frac{\delta_{\pm}}{8\pi}\nonumber\\&=&\frac{1}{4}\left(1-\frac{H_{\pm}}{K}\right)\\&=&\frac{1}{4}\left(1-\frac{1\pm 2mA+ q^2 A^2/\eta}{K}\right),
\end{eqnarray}
where $\delta_{\pm}$ represents conical deficits at the north and south poles, $H_{\pm}$ stands for $H(\theta=0)$ and $H(\theta=\pi)$.  We can also define \cite{Gregory:2019dtq}
\begin{equation}\label{deltax}
\Delta \equiv 1-2(\mu_{+}+\mu_{-}) = \frac{A^2 q^2+\eta}{\eta K},
\end{equation}
\begin{equation}\label{cx}
C \equiv \frac{\mu_{-}-\mu_{+}}{\Delta} =  \frac{m A \eta}{A^2 q^2+\eta},
\end{equation}
which respectively are average and differential conical deficits 
for the charged $f(R)$ accelerating black hole. We note that
as $\eta\to 0$ we obtain a large average deficit $\Delta$ and a vanishingly small  differential  deficit $C$; conversely for large $\eta$ both $\Delta$ and $C$ approach their values in the charge neutral case. Since both $\mu_{\pm}$ are dimensionless, these quantities, along with their thermodynamic conjugates $\lambda_{\pm}$ \cite{Appels:2017xoe},  do
not enter the Smarr relation (\ref{smarr}). There is a direct but rather cumbersome method to
obtain these quantities from the first law \cite{Anabalon:2018qfv}. Rather than proceeding in this manner, we can instead write the mass entirely in terms of
the extrinsic thermodynamic parameters, obtaining
\begin{equation}\label{mas}
\begin{aligned}
M^{2}=\frac{\Delta  \eta  S}{4 \pi }&\left[\left(1+\frac{8 P S}{3 \Delta  \eta }+\frac{\pi   Q^2}{\Delta  S}\right)^2\right.\\ &\left.-\frac{3 \Delta  \eta  C^2}{2PS} \left(\frac{8 P S}{3 \Delta  \eta }+1\right)\right].
\end{aligned}
\end{equation}

From this we can compute the temperature, electric potential and thermodynamic volume as
 \begin{equation}
\begin{aligned}
T=&\left(\frac{\partial M}{\partial S}\right)_{Q,P,\mu_{\pm}}\\=&\frac{1}{24 \pi  \Delta  \eta  M S^2}\left[64 P^2 S^4+16 \eta  P S^2 \left(\pi  Q^2+2 \Delta  S\right)\right.\\ &\left.-3 \eta ^2 \left(\pi ^2  Q^4+\Delta ^2 S^2 \left(4 C ^2-1\right)\right)\right],
\end{aligned}
\end{equation}
\begin{equation}
\Phi=\left(\frac{\partial M}{\partial Q}\right)_{S,P,\mu_{\pm}}=\frac{  Q \left(8 P S^2+3 \eta  \left(\pi  Q^2+\Delta  S\right)\right)}{6 \Delta  M S},
\end{equation}
\begin{equation}
\begin{aligned}\label{masvol}
V=&\left(\frac{\partial M}{\partial P}\right)_{S,Q,\mu_{\pm}}\\=&\frac{27 \Delta ^3 \eta ^3 C ^2+256 P^3 S^3+96 \eta  P^2 S \left(\pi  Q^2+\Delta  S\right)}{144 \pi  \Delta  \eta  M P^2},
\end{aligned}
\end{equation}
which, upon using the expressions \eqref{accent}, \eqref{ele}, and \eqref{pressure} can be shown to agree  with those in (\ref{temp}), (\ref{pot}) and (\ref{vol}) respectively.

The thermodynamic lengths  conjugate to the tensions $\mu_{\pm}$ are therefore straightforwardly obtained
\begin{equation}\label{length}
\begin{aligned}
\lambda_{\pm}=&\left(\frac{\partial M}{\partial \mu_{\pm}}\right)_{S,Q,P}\\=&\frac{1}{72 \pi  \Delta ^2 \eta  M P S}\left[128 P^3 S^4+96 \pi  \eta P^2 Q^2 S^2\right.\\  &\left.-18 \eta ^2 P \left(\Delta ^2 S^2 (1\mp 2 C )^2-\pi ^2  Q^4\right)\right. \\&\left.\pm 27 \Delta ^3 \eta ^3 S C\right].
\end{aligned}
\end{equation} 
and it is then easy to check that the first law 
\begin{equation}\label{flaweta}
d M=T dS +\Phi d Q+\lambda_{+}\ d\mu_+ +\lambda_- d\mu_- +V d P 
\end{equation}
is satisfied.

The key advantage of the formulation \eqref{mas} is that it 
elucidates the chemical structure of the accelerating black hole \cite{Gregory:2019dtq}.  Indeed, using \eqref{mas} and \eqref{masvol} we obtain a generalization of the Reverse Isoperimetric Inequality \cite{Cvetic:2010jb,Cong:2019bud,Mann:2018jzt,Sinamuli:2015drn}
\begin{equation}
\begin{aligned}\label{riso2}
\left(\frac{3V}{4\pi} \right)^2 \geq \frac{1}{\eta\Delta}  \left(\frac{S}{\pi} \right)^3=\frac{\eta^2}{\Delta}\left(\frac{\mathcal{A}}{4\pi} \right)^3
\end{aligned}
\end{equation}
for accelerating $f(R)$ black holes, where $\mathcal{A}$ is the horizon area.  The equality is
saturated if $C=0$, and reduces to that in Einstein gravity  \cite{Gregory:2019dtq} if $\eta=1$.
A larger deficit yields a smaller $\Delta$, and therefore a smaller entropy with respect to the volume,
tending to make the black holes sub-entropic.  However a larger value of $\eta$ works in a contrary manner, allowing for a larger entropy (but smaller area!)
relative to the volume, tending to make the black holes super-entropic \cite{Hennigar:2014cfa,Hennigar:2015cja}.

We can confirm our result by calculating the action with Gibbons-Hawking boundary term and  boundary counterterms \cite{Emparan:1999pm,Mann:1999pc,Das:2000cu,Guarnizo:2010xr}
\begin{equation}\label{acacc}
\begin{aligned}
I=&I_{\text{bulk}}+I_{\text{surf}}+I_{\text{ct}}\\=&\frac{1}{16\pi}\int_{M}d^{4}x\sqrt{-g}\left[ R+f(R)-F^2\right]
\\&+\frac{1}{8\pi}\int_{\partial M}d^{3}x\sqrt{-h}\eta\mathcal{K} \\
&- \frac{1}{8\pi}\int_{\partial M}d^{3}x\sqrt{-h}\eta \mathcal{H}(l,\mathcal{R},\nabla\mathcal{R}),
\end{aligned}
\end{equation}
where
\begin{equation}
\mathcal{H}(l,\mathcal{R},\nabla\mathcal{R})=\frac{2}{l}+\frac{l}{2}\mathcal{R}-\frac{l^3}{2}\left(\mathcal{R}_{ab}\mathcal{R}^{ab}-\frac{3}{8}\mathcal{R}^2 \right)+\cdots.
\end{equation}
In the bulk term, one should note that $R+f(R)=-6\eta/l^{2}$, $\mathcal{K}$ is the extrinsic curvature scalar of the boundary surface, $\mathcal{R}$ is the intrinsic curvature of the conformal boundary described by the induced metric $h_{ij}$. One should note that the boundary here is the conformal boundary. After calculation, one can obtain the Gibbs free energy $G$ as
\begin{equation}\label{Gibbs}
\begin{aligned}
G=&\frac{I}{\beta}\\
=&-\frac{\eta  r_+^3}{2 \alpha  K l^2 \left(1-A^2 r_+^2\right){}^2}-\frac{ q^2}{2 \alpha  K r_+}-\frac{ \eta  m}{2 \alpha  K}\\&+\frac{  \eta  m \left(1-A^2 l^2\right)}{\alpha  K}-\frac{ m A^4 l^2 q^2}{\alpha K}\\
=&M-TS-\Phi Q.
\end{aligned}
\end{equation}
We have neglected the boundary term
\begin{equation}\label{Fbndy}
I^{\text{em}}_b =\frac{1}{4\pi}\int_{\partial\mathcal{M}}d^3 y\sqrt{|h|}n_a F^{ab}B_b
\end{equation}
corresponding to the electricmagnetic field in the action (\ref{acacc}), which means that the system's electric potential is fixed but the electric charge is variable \cite{Brown:2018bms,Goto:2018iay,Liu:2019smx,Jiang:2019qea}.

\section{Phase transition of the slowly accelerating $f(R)$ A\lowercase{d}S black hole}\label{secth}

\subsection{Parameter space}

We will discuss permitted range of the parameters $m,~A,~q,~l,~\eta$ for the slowly accelerating $f(R)$ AdS black hole. Following convention in \cite{Abbasvandi:2018vsh}, we define 
\begin{equation}
\tilde{m}=mA,~\tilde{q}=qA,~\tilde{A}=Al,~\tilde{r}=\frac{r}{l},
\end{equation}
which are dimensionless quantites, then $N(r)$ and $H(\theta)$ can be written as
\begin{equation}
N(\tilde{r})=(1-\tilde{A}^{2}\tilde{r}^{2})\left(1-\frac{2\tilde{m}}{\tilde{A}\tilde{r}}+\frac{\tilde{q}^{2}}{\eta\tilde{A}^{2}\tilde{r}^{2}}\right)+\tilde{r}^{2},
\end{equation}
\begin{equation}
H(\theta)=1+2\tilde{m}\cos\theta+\frac{\tilde{q}^{2}}{\eta}\cos^{2}\theta.
\end{equation}

First, we must ensure the existence of an outer event horizon $\tilde{r}_{+}$ such that
\begin{equation}
N(\tilde{r}_{+})|_{0<\tilde{r}_{+}<1/\tilde{A}}=0,
\end{equation}
with $\tilde{r}_{+}=r_{+}/l$, and the positivity of the blackening factor in the region between the outer horizon and the conformal boundary, which can be expressed as
\begin{equation}\label{co2}
N(\tilde{r})|_{\tilde{r}_{+}<\tilde{r}<1/\tilde{A}}>0.
\end{equation}
The condition (\ref{co2}) can be assured if we have
\begin{equation}\label{ps1}
N^{\prime}(\tilde{r}_{+})|_{0<\tilde{r}_{+}<1/\tilde{A}}>0=N(\tilde{r}_{+})|_{0<\tilde{r}_{+}<1/\tilde{A}}.
\end{equation}

Second, we should make sure
\begin{equation}
H(\theta)|_{-\pi\leqslant\theta\leqslant\pi}>0,
\end{equation}
which is equivalent to let
\begin{equation}\label{ps2}
\tilde{m}<\sqrt{\frac{\tilde{q}^{2}}{\eta}}\varTheta\left(\frac{\tilde{q}^{2}}{\eta}-1\right)+\frac{1}{2}\left(1+\frac{\tilde{q}^{2}}{\eta}\right)\varTheta\left(1-\frac{\tilde{q}^{2}}{\eta}\right),
\end{equation}
where $\varTheta$ is the unit step function and $\varTheta(0)=1/2$.

Third, we must eliminate the acceleration horizon. The extreme condition for the emergence of the acceleration horizon is
\begin{equation}\label{ps3}
N(\tilde{r})|_{\tilde{r}=-1/(\tilde{A}\cos\theta)}=N^{\prime}(\tilde{r})|_{\tilde{r}=-1/(\tilde{A}\cos\theta)}=0,
\end{equation}
from which we can get $\tilde{m}=\tilde{m}(\tilde{q},\theta,\eta)$ and $\tilde{A}=\tilde{A}(\tilde{q},\theta,\eta)$.

Moreover, one should note that
\begin{equation}\label{res2}
1-A^2 l^2 \Xi>0
\end{equation}
in (\ref{alpha}), which is equivalent to
\begin{equation}\label{ps4}
\tilde{A}<\sqrt{\frac{\eta}{\eta+\tilde{q}^{2}}}.
\end{equation}

\begin{figure}[t]
    \centering
    \includegraphics[width=3.5in]{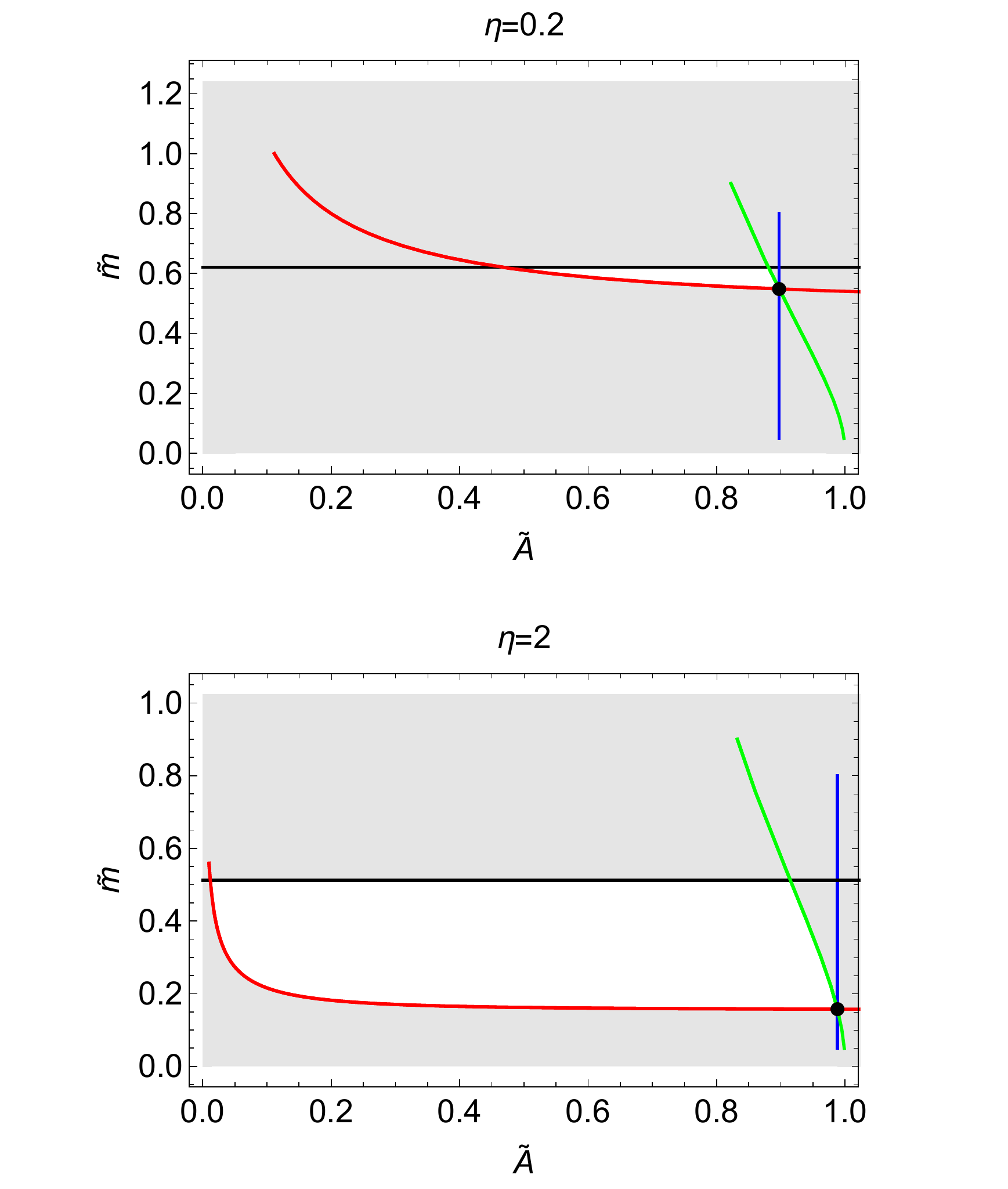}
    \caption{The permitted parameter spaces for $\eta=0.2, 2$ and $\tilde{q}=11/50$ are the white enclosed regions. The red, black, green, and blue  lines respectively correspond to conditions (\ref{ps1}), (\ref{ps2}), (\ref{ps3}) and (\ref{ps4}).}
    \label{para}
 \end{figure}

The parameter space has been shown in Fig. \ref{para} with white enclosed regions. We have chosen two different values for characteristic quantity $\eta$ of the charged accelerating black hole in $f(R)$ gravity. It is evident that a larger $\eta$ is endowed with a larger physically reasonable parameter space.  There is a point where the blue, green and red curves intersect (we name it point $X$, as the convention in \cite{Abbasvandi:2018vsh}), which can be solved for, yielding
\begin{equation}\label{pointx}
\left(\tilde{m}_{X}, \tilde{A}_{X}\right)=\left(\frac{\sqrt{\tilde{q}^{2}(\eta+\tilde{q}^{2})}}{\eta}, \sqrt{\frac{\eta}{\eta+\tilde{q}^{2}}}\right).
\end{equation}
Combining the expressions (\ref{ps1}), (\ref{ps2}), (\ref{ps3}), (\ref{ps4}) and (\ref{pointx}) with the figures, when $\eta\to\infty$, we obtain
\begin{equation}
|\tilde{q}|<\tilde{m}<\frac{1}{2}\left(1+\frac{\tilde{q}^2}{\eta}\right)\to\frac{1}{2},
\end{equation}
and
\begin{equation}
0<\tilde{A}<1,
\end{equation}
whereas for $\eta\to 0$ (note that $\eta>0$ for charged accelerating $f(R)$ black hole), we have
\begin{equation}\label{pscon}
|\tilde{q}|\sqrt{1+\frac{\tilde{q}^{2}}{\eta}}<\tilde{m}<\frac{|\tilde{q}|}{\sqrt{\eta}},
\end{equation}
and
\begin{equation}
0<\tilde{A}<\tilde{A}_X\to 0^+ .
\end{equation}
(\ref{pscon}) demands that $|\tilde{q}|<1$.

When all curves intersect into one point (we can name it point $E$), which makes the permitted region to be none, we can obtain the coordinate of the extreme point as
\begin{equation}
\left(\tilde{q}_{E}, \tilde{m}_{E}, \tilde{A}_{E}\right)=\left(\sqrt{\frac{\eta}{3}}, \frac{2}{3}, \frac{\sqrt{3}}{2}\right)\, .
\end{equation}
 When $\eta=1$, the situation reduces to that studied previously \cite{Abbasvandi:2018vsh} for
 charged accelerating black holes in Einstein gravity.

\begin{figure}[t]
    \centering
    \includegraphics[width=3in]{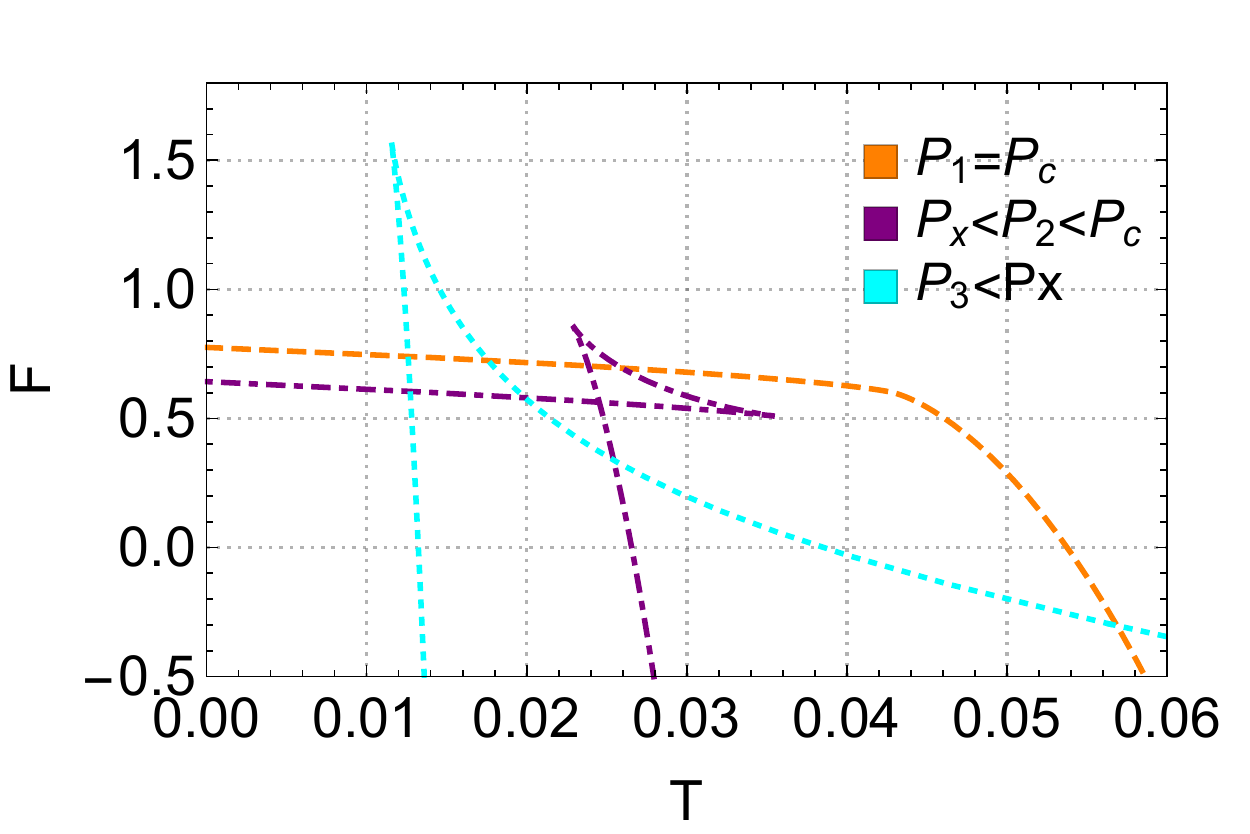}
    \caption{The free energy in terms of the temperature in the canonical ensemble with  $\eta=0.8, \mu_{+}=0, \mu_{-}=0.05, Q=0.8, P_{1}=P_{c}=1/(96\pi), P_{2}=P_{1}/4=25 P_{X}/9, P_{3}=P_{1}/16=25 P_{X}/36$.}
        \label{ft}
 \end{figure}

\subsection{Phase transitions}

We now consider the phase behaviour  of these slowly accelerating black holes.  We begin with the canonical ensemble, where the electric charge of the black hole is fixed. The Helmholtz free energy for this ensemble, which can be obtained by including \eqref{Fbndy} in the computation of the action, is
\begin{equation}
F=M-TS.
\end{equation}

 \begin{figure}[h!]
    \centering
   ~~~~~\includegraphics[width=3.1in]{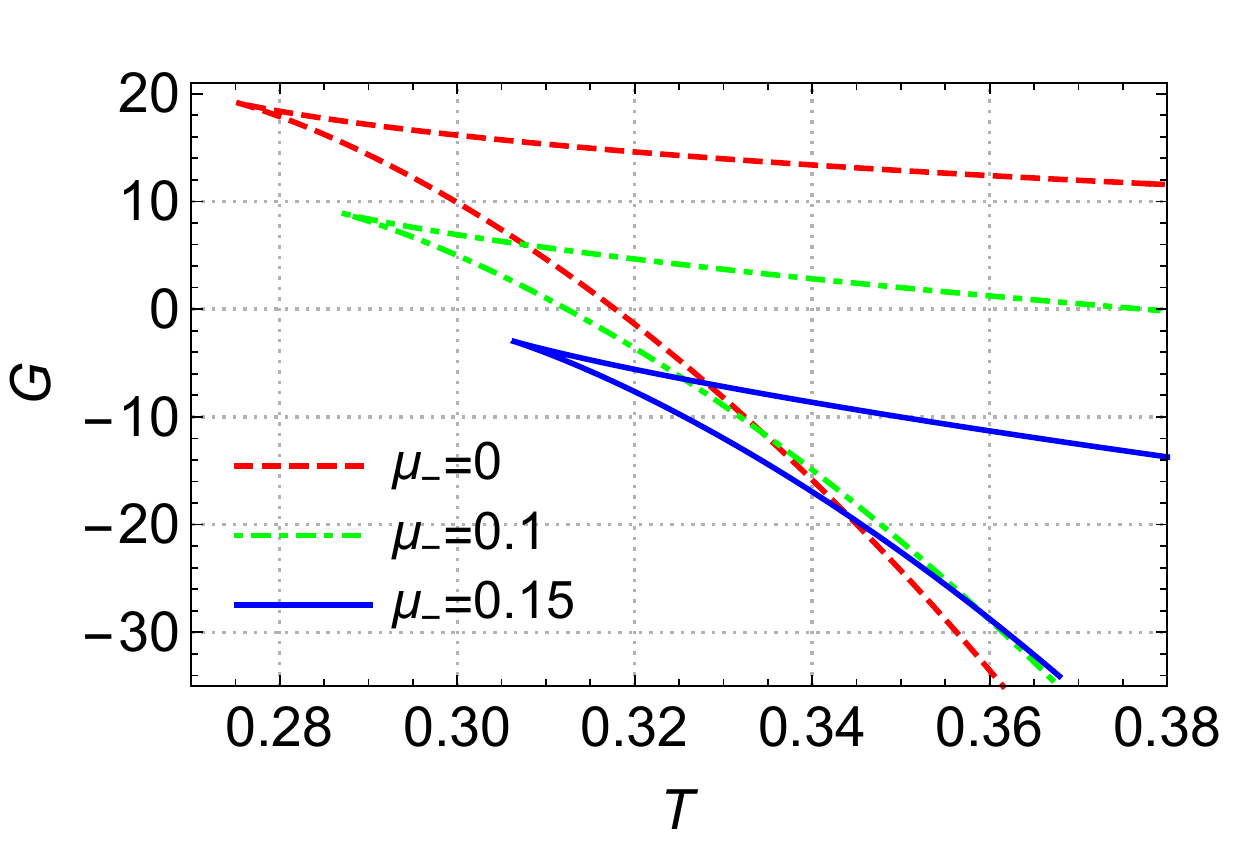}\\
     \includegraphics[width=3.4in]{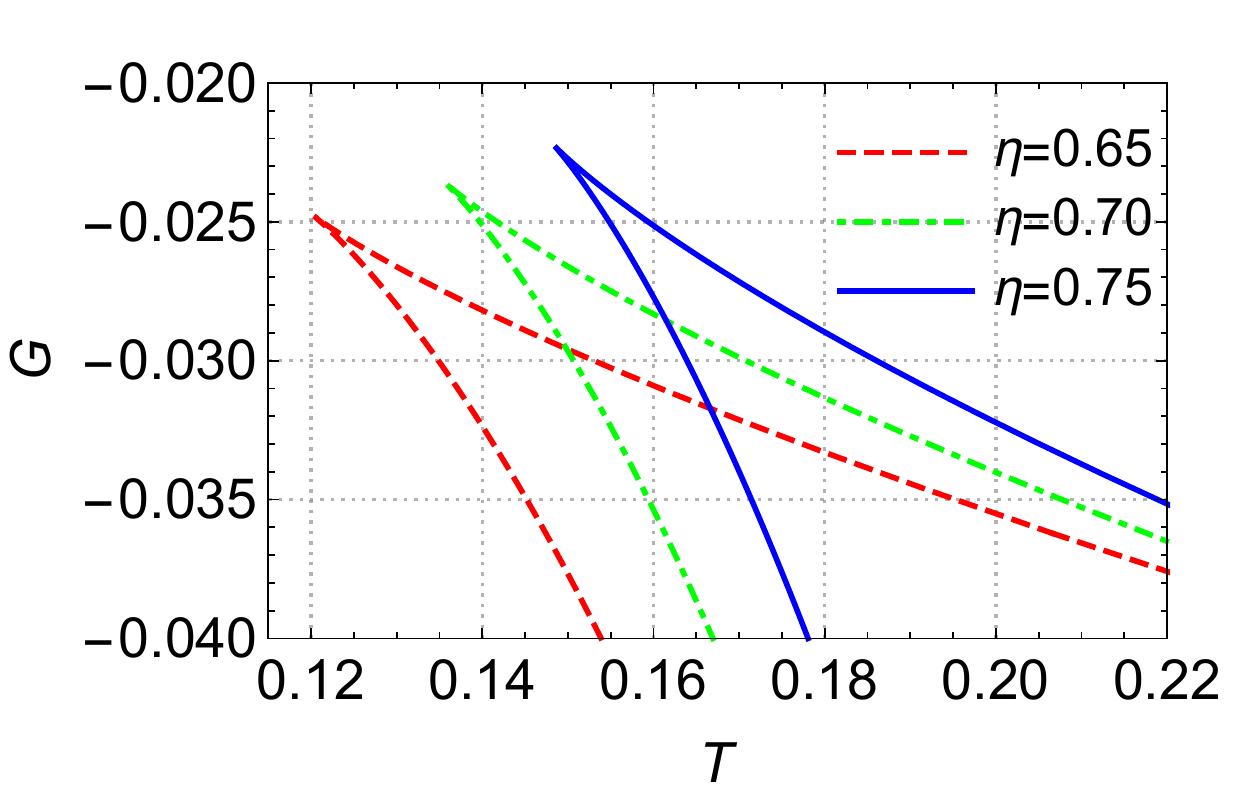}\\
     ~~\includegraphics[width=3.28in]{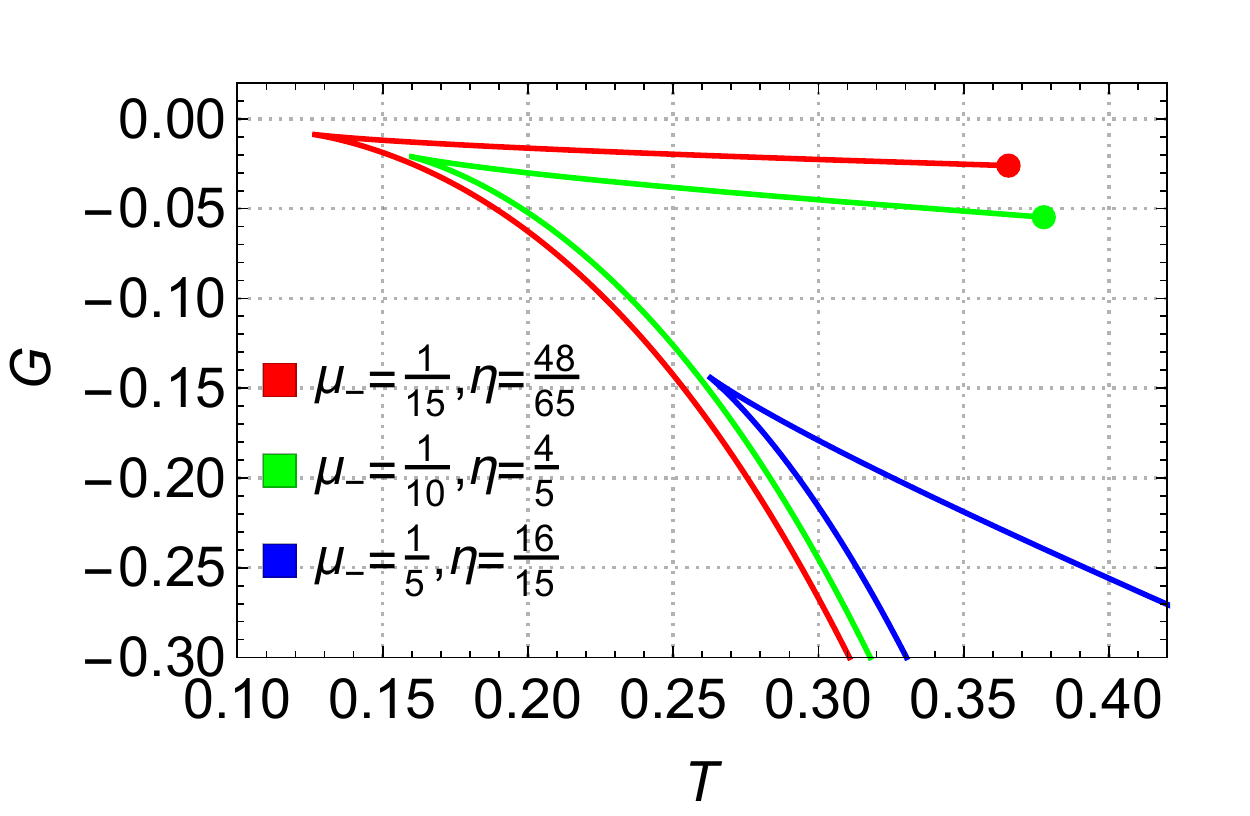} 
    \caption{ The Gibbs free energy $G$ in terms of the temperature $T$ for the charged slowly accelerating $f(R)$  AdS black hole with $\Phi=0.8, \mu_+ =0, P=3/(8\pi)$. In the top diagram $\eta=200$ and $\sqrt{\Delta\eta}>\Phi$; in the middle diagram $\mu_{-} = 0.1$ and $\sqrt{\Delta\eta}<\Phi$; in the bottom diagram $\sqrt{\Delta\eta}=\Phi$. Different values of $\Phi, \mu_+$ and $P$ do not quantitatively affect our results. To a numerical precision of  0.01 we find in the top diagram that the upper branches terminate at $T=27.87,~19.64$ for $\mu_- =0.1,~0.15$ respectively, and that the terminal temperature decreases for increasing $\mu_-$. In the middle diagram, the upper branches   terminate at $T=0.30,~0.33,~0.36$ for $\eta=0.65,~0.70,~0.75$ respectively, and  the terminal temperature increases for increasing $\eta$. In the bottom diagram, the upper branches   terminate for values of $(T,G)$ at
(0.37, -0.026), (0.38, -0.055), (0.70, -0.465) individually.}
        \label{goft}
 \end{figure}

Without loss of generality \cite{Abbasvandi:2018vsh}, we  choose $K=\Xi+2\tilde{m}$, so that   $\mu_{+}=0$.  The conical deficit is located entirely along the south  pole axis $\theta=\pi$, with $\mu_{-}=\tilde{m}/K$. According to (\ref{ele}), (\ref{pressure}), (\ref{deltax}), (\ref{cx}) and (\ref{pointx}), the pressure corresponding to the point $X$ is
\begin{equation}
\begin{aligned}
P_{X}=&\frac{3\tilde{m}^{2}}{8\pi Q^{2}\left(1+\tilde{q}^{2}/\eta+2\tilde{m}\right)}\\=&\frac{3\mu_{-}^{2}}{8\pi Q^{2}}\\=&\frac{3 (2 C \Delta -\Delta +1)^2}{128 \pi  Q^2},
\end{aligned}
\end{equation}
from which we can see that the parameter $\eta$ does not have a qualitatively different effect on the pressure of the intersection point $X$. As a result, when considering phase structure of the slowly accelerating $f(R)$ AdS black hole, we see that it will be qualitatively similar to that
in Einstein-AdS gravity \cite{Abbasvandi:2018vsh}.  

In other words, a slowly accelerating $f(R)$ AdS black hole will experience a snapping swallowtail process when the pressure of the black hole decreases  below $P_{X}$, and the black hole will experience zeroth order, first order, second order and reentrant phase transitions \cite{Abbasvandi:2018vsh}. We depict the Helmholtz free energy in terms of the temperature in Fig. \ref{ft}, from which we can see that the phase transition of the slowly accelerating $f(R)$ AdS black hole is van-der Waals like for $P>P_{X}$, and the branch corresponding to the small black hole with low temperature has snapped for $P<P_{X}$.

Turning to  the grand canonical ensemble, where the electric potential of the black hole is fixed, we 
plot the  Gibbs free energy  \eqref{Gibbs} in terms of the temperature to see the phase structure. Results are shown in  Fig. \ref{goft} for fixed $\eta$ (top diagram), fixed $\mu_{-}$ (middle diagram) and fixed $\sqrt{\Delta\eta}$ (bottom diagram).

Setting  $C=0$  but $\Delta\neq 1$  (i.e.,  $\mu_- =\mu_+ \neq 0$) we obtain
the zero-acceleration $f(R)$ black hole threaded by a cosmic string.  In this case, by analyzing the derivative of the temperature   with respect to the entropy   
\begin{equation}\label{speca}
T^\prime (S)=\frac{\Delta \eta  \left(8 P S+\Phi ^2-\Delta \eta \right)}{8 \sqrt{\pi } (S\Delta \eta)^{3/2}},
\end{equation}
we find that $\sqrt{\Delta\eta}>\Phi$ yields curves similar to those in the top diagram in Fig. \ref{goft}, 
whereas $\sqrt{\Delta\eta}\leqslant\Phi$ yields   the  Gibbs free energy monotonically decreasing  with temperature (not illustrated here), a phenomenon noted previously for the zero-acceleration case without a cosmic string \cite{Li:2016wzx,Zhang:2018rlv} .  For $\Delta=1,~$ $C=0$ (i.e., $\mu_- =\mu_+ = 0$) and
 $\sqrt{\eta}>\Phi$ we obtain the uppermost (red) curve shown in the top diagram in Fig. \ref{goft},  
for which a standard Hawking-Page transition occurs at $G=0$.     We do not have an interpretation of what the transition at $G=0$ means for  $\mu_{-} >0$.

 For the charged slowly accelerating $f(R)$ black hole, we find that $T^{\prime}(S)=0$ always has a solution  for $S>0$. As a result  Hawking-Page(-like) curves for the Gibbs free energy are always present. From Fig. \ref{goft}, we can see that the lowest temperature attainable by  a charged slowly accelerating $f(R)$ black hole increases for increasing $\mu_-$ and $\eta$.    We also find that there exist curves that  at low temperatures have $G>0$ on portions of both branches, whereas at high temperatures $G<0$ on both branches.    The lower branch corresponds to stable black holes when $G<0$.  The upper branches terminate at some finite large temperature that is not visible in the diagrams (except for the two terminal points shown in the bottom diagram of Fig. \ref{goft}), and which increase for decreasing $\mu_-$ and for increasing $\eta$.

\section{Summary}\label{secf}

 We have obtained the first accelerating black hole in modified gravity ($f(R)$ gravity), and have investigated its thermodynamics, computing its mass, entropy, temperature, electric charge, electric potential, and tensions.  We find a number of similarities to the charged accelerating case
studied previously in Einstein gravity \cite{Abbasvandi:2018vsh}. The structure of its thermodynamic phase transitions  are similar to those in Einstein gravity, and the `snapping swallowtail' phenomenon
is exhibited. 

 However  we also find a number of distinctions due to the parameter $\eta$ that arise because
the mass and entropy are both linearly dependent on this quantity.  This means that it cannot be
scaled out of the charge parameter, despite superficial appearances to the contrary.  We also obtain
a generalization of the reverse isoperimetric inequality \eqref{riso2}, suitable for $f(R)$ gravity. 
While the dependence on the conical deficit is (as expected) the same as in Einstein gravity
\cite{Gregory:2019dtq}, we find that increasing $\eta$ tends to allow the black holes to have more area relative to their thermodynamic volume, tending to make them super-entropic.  
Furthermore, the allowed region of parameter space governing the phase transitions becomes $\eta$-dependent, broadening the range of possibilities.  We found that for increasing $\eta$ the accessible  parameter space is enlarged. Specifically, we gave the upper and lower bounds for the parameters $\tilde{m}$ and $\tilde{A}$ that characterize the parameter space when $\eta$ varies between $0$ and $\infty$.  We have analyzed the phase structures of the charged slowly accelerating $f(R)$ AdS black hole both in the canonical ensemble and grand canonical ensemble. In the fixed-charge ensemble, we found van der Waals phase transitions and snapping swallowtail behaviour, similar to that in Einstein gravity\cite{Abbasvandi:2018vsh}. In the ensemble where the electric potential is fixed, we found that the lowest temperature the black hole can attain increases for increasing $\mu_-$ and $\eta$.  We also found that there are always two branches in the $G-T$ diagram for $\sqrt{\Delta\eta}>\Phi$, $\sqrt{\Delta\eta}=\Phi$ and even $\sqrt{\Delta\eta}<\Phi$, where the upper one terminates at a high temperature. The terminated temperature decreases for increasing $\mu_-$ and decreasing $\eta$.

$f(R)$ gravity enables us to study the inflation and accelerated expansion epochs of  our universe. It can uniformly explain the early-time inflation as well as the dark energy epoch \cite{Nojiri:2010wj}. It has been testified by many applications in cosmology and astronomy \cite{Capozziello:2011et}, it has also been constrained  by realistic FRW cosmology \cite{Nojiri:2006gh} and stability consideration \cite{Sawicki:2007tf}. $f(R)$ gravity is such a well-behaved modified theory that it can show the evolution of the universe and it can also give correct cosmological predictions \cite{Nashed:2019uyi,Dunsby:2010wg}. It has been shown that $f(R)$ theory without cosmological constant can reproduce the behaviour of  Einstein-Hilbert action with cosmological constant \cite{delaCruzDombriz:2006fj}. 

The charged accelerating black hole in $f(R)$ gravity should provide further insight into $f(R)$ gravity. The role of $\eta$ needs to be better understood.  For example, for $\eta > 1$, will
these black holes exhibit   thermodynamic instabilities similar to those found recently \cite{Johnson:2019mdp} for other super-entropic black holes?  
It is easy to extend our solution for the charged accelerating black hole in $f(R)$ gravity to   one with rotation, and it is possible that interesting $\eta$-dependent phenomena may be present in this case.
Likewise, the possibility that $\eta$ itself could be a thermodynamic parameter merits investigation.
Work on these topics is in progress.
\bigskip

\section*{Acknowledgements}
The authors would like to thank Wan Cong, Niloofar Abbasvandi and Jie Jiang for helpful discussions. This work was supported in part by the Natural Sciences and Engineering Research Council of Canada.

\end{document}